\documentclass[11pt]{article}
\fontfamily{times}
\usepackage{graphicx}
\usepackage{geometry, amsmath}
\usepackage{bm}
\newtheorem{thm}{Theorem}

\newtheorem{lemma}{Lemma}

\newcommand{\ben}{\begin{enumerate}}
\newcommand{\een}{\end{enumerate}}
\newcommand{\beq}{\begin{eqnarray}}
\newcommand{\eeq}{\end{eqnarray}}
\newcommand{\beqn}{\begin{eqnarray*}}
\newcommand{\eeqn}{\end{eqnarray*}}
\newcommand{\e}{\varepsilon}

\newcommand{\be}{\begin{equation}}
\newcommand{\ee}{\end{equation}}
\font\Bbb=msbm10

\def\sk1{\vskip 10pt}

\def\Bbb#1{I\!\! #1}

\def\y{{\bold y}}
\def\X{{\bold X}}
\def\x{{\bold x}}

\def\f{{\bold f}}
\def\e{{\bold e}}
\def\t{{\tt t}}

\def\b{{\bold b}}

\def\z{{\bold z}}

\def\w{{\bold w}}

\def\c{{\bold c}}

\def\b0{{\bold 0}}

\def\bold{\bf}

\def\t{{\bold t}}
\def\f{{\bold f}}

\def\BSigma{\boldsymbol{\Sigma}}
\def\thn{\boldsymbol{\hat\theta}_n}
\def\ths{\boldsymbol{\hat\theta}_n^*}
\def\thz{\boldsymbol{\theta}_0}
\def\tth{\boldsymbol{\tilde{\theta}}_n}
\def\tht{\boldsymbol{\theta}}
\def\bphi{\boldsymbol{\phi}}

\def\ssum{\sum_{i=1}^{n}}

\font\absmall=cmr6 at 6 pt
\def\Cal{\cal}
\def\l{\left[}
\def\r{\right]}
\def\rr{\right)}
\def\ll{\left(}
\providecommand{\keywords}[1]{\textbf{\textit{Keywords: }} #1}

\newcommand{\qed}{\hspace*{\fill}Q.E.D.}  

\title{Recycled Least Squares Estimation in Nonlinear Regression}
\author{Ben Boukai\thanks{%
Email: bboukai@iupui.edu } \ and Yue Zhang\thanks{%
Email: yz65@umail.iu.edu}\\ Department of Mathematical Sciences, IUPUI\\
Indianapolis, Indiana, 46202 }

\begin{document}
\maketitle


\begin{abstract}
\noindent We consider a re-sampling  scheme for parameters' estimates in nonlinear regression models. We provide an estimation procedure which {\it recycles}, via random weighting, the relevant parameters estimates to construct consistent estimates of the sampling distribution of the various estimates. We establish the asymptotic normality of the resampled estimates and demonstrate the applicability of the {\it recycling} approach in a small   
simulation study and via example. 
\end{abstract}


\bigskip

\setlength{\parindent}{0pt}

\section{Introduction} 
One of the most commonly used approaches for estimation of the parameters in nonlinear regression models is that of the least squares method.  However,  these least squares estimators do not generally exhibit tractable nor optimal finite-sample properties, largely due to nonlinearity of the regression function.  Consequently,  statistical inference in these cases relies primarily on either, asymptotic results and/or simulations. Specifically,  consider the following regression model which specifies the relationship between the observations on a response $Y$ and the corresponding $\x=(x_1, x_2, \dots, x_p)^\t\in \Bbb{R}^p$, with $p>1$, via a nonlinear function $f(\cdot)$ as
\be\label{1}
\ y _i=f(\x_i; \, \tht)+\epsilon_{i},   \ \ \ \     i=1,\dots, n 
\ee
where $ \x_i $ is the $ i $th fixed input which gives rise to the observation $ y_i $. $ \tht=(\theta_1, \theta_2, \dots, \theta_p)^\t$ is the unknown $p\times 1$ parameter from a compact parameter space $ \Theta\subset \Bbb{R}^p.$ Here 
$$
f_i(\tht):= f(\x_i; \, \tht) ,   \ \ \ \     i=1,\dots, n 
$$ 
are assumed to be continuous functions in $ \tht\in\Theta $ and $\epsilon_1, \epsilon_2, \dots, \epsilon_n$ are independent and identically distributed (i.i.d.) random error terms with mean 0 and unknown variance $\sigma^2$.   We denote 
by $\thz$ the true, though unknown, value of $\tht$ in (\ref{1}).  We assume throughout that $\thz\in int(\Theta)$. The ($p$-variate) minimizer, $\thn$,  of the residual sum of squares  
\be\label{2}
Q_n(\tht):=\sum_{i=1}^{n}(y_i-f_i(\tht))^2,
\ee
is the least square estimator (LSE) of $\thz$. The existence and strong consistency of $ \thn$ as an estimator of  $\thz$  was established by Jennrich (1969) under the rather general conditions of model (\ref{1}) and {\it Condition J} below. Under some additional conditions (see {\it Assumption A}, below), Wu (1981) established the asymptotic normality of $ \thn$, so that
\be\label{3}
\sqrt{n}(\thn-\thz)\Rightarrow {\cal N}_p(\boldsymbol{0},\sigma^2\BSigma_0), 
\ee
where $\BSigma_0$ is some $p\times p$-positive definite matrix (see {\it Assumption A}, below) and ${\cal N}_p$ denotes the $p-$variate normal distribution.

 In this paper, we are interested in the re-sampled version of the LSE $\thn$ and  its properties. The implementation of various  bootstrapping techniques in the context of linear regression models has received much attention in the literature. Various approaches, such as the `naive', the `residual', the `pair' and the `wild' bootstarp have been thoroughly  studied under various conditions (see, Efron (1979), Efron and Tibshirani (1986), Wu (1986)).  However, here we focus attention on a resampling approached for the nonlinear regression model in (\ref{1}) based on random weighting technique (see for example Zheng and Tu (1988), Mason and Newton (1992), or Chatterjee and Bose (2005)). 

In Section 2 we review and provide for completeness some of the known results on LSE in the non-linear regression context. In Section 3 we discuss the random weights we employ and describe the  re-sampling procedure we propose for estimating the sampling distribution of $\thn$. We term the resulting estimates,  $\ths$, as obtained using  this random-weighting scheme, as a {\it recycled} estimate of $\thz$. We study the conditions for its  strong consistency and the asymptotic normality of the {\it recycled}  version $\ths$ of $\thn$. In Section 4, we provide the results of a simulation study along with a numerical illustration. For the random weighting technique, it is also of interest to compare their relative performances for different choices of random weights and their distributions, which are also provided in Section 4 along with a closing discussion. Section 5 is dedicated to technical details and proofs.

\section{On the LS Estimation}

Jennrich (1969) provided the existence and strong consistency  of $ \thn$, the LSE of $\thz$ for the nonlinear regression model in (\ref{1}) as the minimizer of (\ref{2}),   under the following general condition :

\noindent {\it \underline{Condition J:} $D_n(\tht,\tht^\prime)\equiv {\frac{1}{n}}\sum_{i=1}^{n} (f_i(\tht)-f_i(\tht^\prime))^2 \longrightarrow D(\tht,\tht^\prime)$ uniformly $n\to \infty$, 
where $D(\cdot, \, \cdot)$ is a continuous function for any $\tht, \ \tht^\prime\in \Theta$ and  $D(\tht,\thz)=0 $ if and only if $\tht=\thz$. 
}

As is readily available from (\ref{2}), the least squares estimator $\thn$ of $\thz$ is the $p$-simultaneous solution of $\nabla Q_n(\tht):= 2\sum_{i=1}^{n}\bphi_{i}(\tht)=\boldsymbol{0}$, 
with 
\begin{equation}\label{a}
	\bphi_{i}(\tht)=-(y_i-f_i(\tht))\nabla f_i(\tht), \ \ i=1\dots, n,  
\end{equation}
and where $\nabla f_i(\tht)=(h_{i:j}), \ j=1, \dots, p$ is a $p \times 1$ vector of the partial derivatives of $f_i,$ with $h_{i:j}= \partial f_i(\tht)/ \partial \theta_j,  \ i=1, \dots, n$. Similarity we denote by  
$\nabla^2 f_i(\tht)=(a_{i:jk}(\tht)),\ \ j, k=1\dots p$ the $p\times p$ matrix of partial second derivatives of $f_i$, with  $a_{i:jk}(\tht) := \partial^2 f_i(\tht)/ \partial \theta_j\partial\theta_k, \ i=1, \dots, n$.  
Along with this notation, we denote by $\nabla \bphi_{i}(\tht)$, the $p\times p$ matrix,  
\be\label{d}
\nabla \bphi_{i}(\tht)= \nabla f_i(\tht)\nabla f_i(\tht)^\t -(y_i-f_i(\tht))\nabla^2 f_i(\tht).
\ee
Additionally, we set
\be\label{b}
\BSigma^{-1}_n(\tht):= \frac{1}{n}\sum_{i=1}^{n}\nabla f_i(\tht)\nabla f_i(\tht)^\t, 
\ee
so that by (\ref{a}) and (\ref{b}), 
\be\label{e}
\sum_{i=1}^{n}E\left[ \bphi_i (\thz)\cdot (\bphi_i (\thz))^\t\right] \equiv n\sigma^2 \BSigma^{-1}_n(\thz).  
\ee

Wu (1981) assumed that aside from Condition J, the functions $f_i$ and their gradients satisfy the following conditions (which we will also use in the sequel):

\noindent {\it \underline{Assumption A:} The functions $\f(\tht)=\left(f_1(\tht), f_2(\tht), \dots, f_n(\tht)\right)^\t$ are such that
\begin{enumerate}
	\item $\nabla f_i(\tht)$ and $\nabla^2 f_i(\tht)$ exist for all $ \tht $ near $\thz $;
	\item $\BSigma^{-1}_n(\thz) \to \BSigma_0^{-1}$ a $p\times p$ positive definite  matrix, as $n\to \infty$;
	\item $\BSigma^{-1}_n(\tht)\to \BSigma^{-1}_0$, uniformly\ \  as $\ n\to\infty$ and $|\tht-\thz|\to \boldsymbol{0}$;
	\item there exist a $\delta>0$, such that for all $j, k$ \ \ $ \underset{n\to\infty}{\limsup}\frac{1}{n}\sum_{i=1}^{n}\underset{|\tht-\thz|\le\delta}{\sup}\, a^2_{i:jk}(\tht)<\infty$
	\item if, for any pair (j,k), $\sum_{i=1}^{n}\underset{{|\tht-\thz|\le\delta}}{\sup}\, a^2_{i:jk}(\tht)=\infty$, then there exists a constant  $M$ ( independent of $i$) such that 
	\beqn
	\sup_{s\ne t; s,t\in \Theta_\delta}\left|\frac{\partial^2f_i(s) }{\partial s_js_k}-\frac{\partial^2f_i(t) }{\partial t_jt_k}\right|/|s-t|\le M \underset{\tht\in \Theta_\delta}{\sup}|a_{i:jk}(\tht)|< \infty
	\eeqn
	for all $i$, where $\Theta_\delta=\{\tht\in\Theta, |\tht-\thz|\le\delta\}$, and $\delta$ is same as in 4 above.

\end{enumerate}
}
Under the terms of {\it Assumption A}, Wu (1981) has established the asymptotic normality of the LSE, $\thn$, as  is given in (\ref{3}). Furthermore, for any $\c=(c_1, c_2, \dots, c_p)^\t$, with $||\c||=1$, we have again under {\it Assumption A}, that  
\be\label{8}
{\Cal{R}}_{n,c}:= \frac{\sqrt{n}\, \c^\t (\thn-\thz) }{\sqrt{\c^\t (\BSigma_n(\thz))\c}} \Rightarrow {\cal N}(0,\sigma^2), \ as, n\to \infty.
\ee
 
In the next section, we describe the re-sampling scheme we use to obtain the {\it recycled} estimate $\ths$ of $\thn$.  We denote by  ${\Cal{R}}^*_{n,c}$ the corresponding recycled version of  ${\Cal{R}}_{n,c}$ in (\ref{8}). 
Let, $P^*$ denote
the {\it resampled} probability  (conditional on the given sample data) and  set
\be
F_n(u)=P({\Cal{R}}_{n,c}\leq u),  \ \ \text{and}\ \  \ F_n^*(u)=P^*({\Cal{R}}^*_{n,c}\leq u),  \ \ \ \forall u\in {\Bbb{R}},
\ee
to denote the corresponding c.d.f of ${\Cal{R}}_{n,c}$ and ${\Cal{R}}^*_{n,c}$, respectively. We show in particular that under the conditions given in Theorem 2 below, 
$$
\sup_u| F_{n}^*(u)-F_n(u)|\to 0 \ \ \ a.s.  \ \ as \ \ \ n\to\infty. 
$$
 This result allows us to approximate the sampling distribution of $\thn$ based on the sampling distribution of its {\it recycled} version, $\ths$.

\medskip

\section{Recycled Estimation via Random Weighting}

For each $n\geq 1$, we let the random weights, $\w_n=(w_{1:n}, w_{2:n}, \dots, w_{n:n})^\t$, be a vector of exchangeable nonnegative random variables with $E(w_{i:n})=1$ and   $Var(w_{i:n}):= \tau_n^2$,   and let $W_{i}\equiv W_{1:n}=(w_{i:n}-1)/\tau_n$ be the standardized version of $w_{i:n}$, $i=1, \dots, n$.  In addition we also assume,

\noindent {\it \underline{Assumption W:} The underlying distribution of the random weights $\w_n$ satisfies
\begin{enumerate}

  \item For all $n\geq 1$, the random weights $\w_n$ are independent of $(\epsilon_1, \epsilon_2, \dots, \epsilon_n)^\t$;
  \item $\tau^2_n=o(n)$, $E(W_iW_j)=O(n^{-1})$ and $E(W_i^2W_j^2)\to 1$   for all $i\ne j$,  $E(W_i^4)<\infty$ for all $i$.
\end{enumerate}
 }
The following are examples of random weights that satisfy the above conditions in {\it Assumption W}. 
\begin{itemize}
  \item[i)] Multinomial weights, $\w_n \sim{\Cal{M}}ultinomial(n, 1/n, 1/n, \dots, 1/n)$, which reflect  a simple random re-sampling scheme (with replacement) and essentially correspond to
the classical bootstrap of Efron (1979).
\item[ii)] Dirichlet weights, $\w_n\equiv n\times \z_n$ where $\z_n\sim {\Cal{D}}irichlet(\alpha, \alpha, \dots, \alpha)$, with $\alpha>0$ which often refer to as the Bayesian bootstrap (see Rubin (1981), and its variants as in Zheng and
Tu (1988) and Lo (1991)). 
\end{itemize}
We will assume throughout this paper that all the random weights we use in the sequel do satisfy {\it Assumption W}.  With  such random weights $\w_n$ at hand, we define in similarity to (\ref{2}),   the {\it recycled} version  $\ths$ of $\thn$ as the minimizer of the {\it randomly weighted} least squares criterion. 
\be\label{10}
Q_n^*(\boldsymbol{\theta}):=\sum_{i=1}^{n}w_{i:n}(y_i-f_i(\boldsymbol{\theta}))^2,  
\ee
or alternatively as the  $p$-simultaneous solution of $\nabla Q_n^{*}(\boldsymbol{\theta}):=2\sum_{i=1}^{n}w_{i:n}\bphi_{i}(\boldsymbol{\theta})=\boldsymbol{0}$. The next results establishes the storng consistency of the {\it recycled} estimator  $\ths$ for $\thz$. 
  
\begin{thm} Let $\ths$ be the minimizer of $Q_n^*(\boldsymbol{\theta})$ in (\ref{10}) and suppose that $E(\epsilon^4_{i})<\infty$ and that in addition to {\it Condition J},  $d_i(\tht):= f_i(\tht)-f_i(\thz)$ also satisfy 
$$
 \underset{n\to\infty}{\limsup}\frac{1}{n}\sum_{i=1}^{n}\underset{|\tht-\thz|\le\delta}{\sup}\, d^4_{i}(\tht)<\infty,
$$
 Then,   $\ths \to \thz $ a.s., as $n\to \infty$. 
\end{thm}

Once $\ths$ is obtained, the {\it recycled} version ${\Cal{R}}^*_{n,c}$ of ${\Cal{R}}_{n,c}$ in (\ref{8}) can readily be obtained as:
\be
{\Cal{R}}_{n,c}^*:= \frac{\sqrt{n}\, \c^\t(\ths-\thn)}{\tau_n\sqrt{\c^\t (\BSigma_n(\thn))\c}}, \label{rstar}
\ee
where $\BSigma^{-1}_n(\thn)$ is as in (\ref{b}), evaluated at $\thn$.  Below we show that under some additional conditions to those stated above, we may obtain the asymptotic normality  of   ${\Cal{R}}_{n,c}^*$. However, aside from the  strong consistency of $\ths $ we will also need the following additional conditions in order to establish the consistency of the resampling estimator for estimating the sampling distribution of $\thn$. 

\vfill\eject

\noindent {\it \underline{Assumption B: } In addition to Assumption A,  we assume that 
\begin{enumerate}
  
\item $ \underset{n\to\infty}{\limsup}\frac{1}{n}\sum_{i=1}^{n}\underset{|\tht-\thz|\le\delta}{\sup}\, a^4_{i:jk}(\tht)<\infty$, for all $j, k$.  
\item  $\underset{n\to\infty}{\limsup}\frac{1}{n}\sum_{i=1}^{n}\underset{|\tht-\thz|\le\delta}{\sup}\, h^4_{i:j}(\tht)<\infty$, for each $j=1, \dots, p$;

\end{enumerate}
}
\begin{thm}{\label{thm2} }	Let $ \thn $ and $\ths$ be a strongly consistent least squares estimator and the recycled estimator of $\thz$ respectively. Then, under {\it Assumptions A \& B} we have, 
$
{\Cal{R}}^*_{n,c} \Rightarrow  {\Cal{N}}(0, \, \sigma^2),
$
as $n\to \infty$,  and hence,  
$$
\underset{u}{\sup}| F_{n}^*(u)-F_n(u)|\to 0 \ \ \ a.s.
$$
\end{thm}

{\bf Remark:} The technique we use for the proof of Theorem 2 (see Section 5 below) can similarity be used, under a stronger version of {\it Assumption B.2},   to also establish the consistency of the (resampling) variance of $\sqrt{n}(\ths-\thn)/\tau_n$. However, this result will be presented elsewhere.

\section{Implementation and Numerical Results}

It is clear from the above description that the {\it recycled} estimate $\ths$ of $\thz$, using the random weights $\w_n$,  is straightforward to implement. Further, in light of Theorem \ref{thm2}, one can also obtain a finite-sample approximation to the sampling distribution of $\thn$ using the ({\it recycled}) sampling distribution of $\ths$. To that end, generate $B$ independent replications of the random weight $\w_n$, independent of the given sample data $(\y, \X)$, to obtain the $B$ recycled replications, $\boldsymbol{\hat\theta}_n^{*1}, \boldsymbol{\hat\theta}_n^{*2}, \dots, \boldsymbol{\hat\theta}_n^{*B}$. Then obtain the $B$ corresponding recycled replications, ${\Cal{R}}^{*1}_{n,c}, {\Cal{R}}^{*2}_{n,c}, \dots, {\Cal{R}}^{*B}_{n,c}$ of ${\Cal{R}}_{n,c}$ as in (\ref{rstar}). Finally, a finite-sample approximation to the sampling distribution, $F^*_n$  of ${\Cal{R}}^*_{n,c}$ can be obtained by the empirical c.d.f  of the  {\it recycled}  ${\Cal{R}}^{*}_{n,c}$, 
$$
\hat F_{B,n}^*(t)={\frac{1}{B}} \sum_{b=1}^{B}I\left[{\Cal{R}}^{*b}_{n,c}\leq t\right].
$$ 

When the error terms variance, $\sigma^2$, is unknown, we may estimate it by  $\hat \sigma^{2}_n:= Q_n(\thn)/(n-p)$ to obtain the `studentized' version, $\hat {\Cal{R}}^*_{n,c}\equiv {\Cal{R}}^{*}_{n,c}/\hat \sigma_n $, of ${\Cal{R}}^{*}_{n,c}$. The  {\it recycled} estimates  of $\hat \sigma^{2}_n$ can be obtained in a similar manner as $\hat \sigma^{*2}_{n,1}, \hat\sigma^{*2}_{n,2}, \dots, \hat \sigma^{*2}_{n,B}$, where $\hat \sigma^{*2}_{n,b}:=Q_n(\boldsymbol{\hat\theta}_n^{*b})/(n-p)$, for $b=1, 2, \dots, B$.

\subsection{A Simulation Study}
For  the simulation study, we considered two nonlinear regression models with $p=2$ each, defined by the functions\\
\noindent {\it \underline{Model I: }}
\[
f(x; \, \tht)=\theta_1x e^{-\theta_2x} 
\]

\begin{figure}[!h]
	\centering 
	\includegraphics[width=0.37\textwidth]{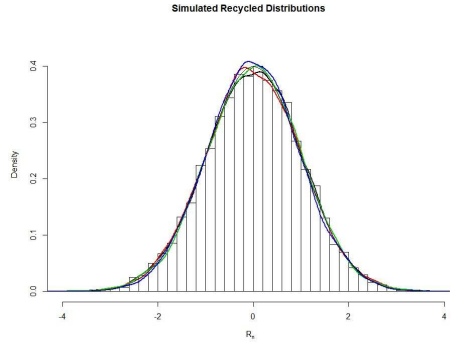}
	\caption{ The Sampling Distribution of ${\Cal{R}}_{n,c}$ ({\tt black}) and the Recycled Distributions using Multinomial ({\tt red}) ,   Dirichlet ({\tt green}) and the Exponential ({\tt blue}), random weights with $n=150$ and $B=10,000$ runs.}
	\label{Fig1}
\end{figure}

\begin{figure}[!h]
	\centering 
	\includegraphics[width=0.37\textwidth]{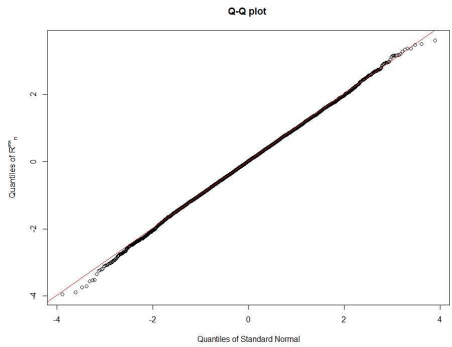}
	\caption{ Normal Probability Plot for the 'Studentized' ${\Cal{R}^*}_{n,c}$ using the Dirichlet random weights. }
	\label{Fig2}
\end{figure}

\noindent {\it \underline{Model II: }}
\[
f(x; \, \tht)=\frac{\theta_1x}{e^{\theta_2}+x}
\]
The fixed values of the independent variable, $x_i, \ i=1, \dots, n$, were selected uniformly from the $[0, 10]$ interval and we assumed that $\epsilon_i\sim {\Cal{N}}(0, 0.25^2)$, $i.i.d.$. In the simulations, the value of the parameter $\tht_0:=(\theta_{10},\theta_{20})$ was set to $(2,0.04)^\prime$ and $(10, 0)^\prime$, corresponding to Model I and Model II,  respectively. For each model, $M=10,000$ {\it recycled} samples were drawn from a sample with varying sample sizes of $n=10, \,30, \,50,\, 80, \,150$. To study the properties of the {\it recycled} estimation procedure, we compared the simulated distribution of ${\Cal{R}}_{n,c}$ to that of the {\it recycled} distribution of $\hat {\Cal{R}}^*_{n,c}$, obtained with $\c=(1/\sqrt{2}, \, 1/\sqrt{2})^\t$. In Tables 1 and 2 below (corresponding to Model I and Model II, respectively),  we present the results as were obtained using random weights $\w_n$,  generated from the standard { (a)}\ {\it Multinomial}, { (b)} {\it Dirichlet} and { (c)} {\it Exponential} distributions.  It can be seen, under either one of these random weighting schemes, the agreement between the simulated and the {\it recycled} distribution of ${\Cal{R}}_{n,c}$ improves, as the sample size increase. This is in agreement with the main results of this paper. To highlight this result, we present in Figure \ref{Fig1} and Figure \ref{Fig2} a comparison of the resulting {\it recycled} sampling distribution of ${\Cal{R}}_{n,c}$ illustrating both, the consistency and the asymptotic normality of the recycled least squares estimate in the case of  Model I.

\begin{table}[!h]
\begin{center}
	{\absmall
		\vskip 2pt
		\begin{tabular}{c c c c c  }	
			\hline\hline
			Sample   & {Simulated} & {Multinomial} & {Dirichlet} & {Exponential }\\
			Size  & {Dist} & {Weights} & {Weights} & {Weights }\\
			\hline
			10               & 0.0037 &-0.2057 &-0.1276& -0.1126 \\
			{\tiny{ S.E.}}   & 1.1441 &0.8529  &0.7193 &0.6534\\ 
			$\hat\sigma$     & 0.2427 &0.3022  &0.2744 &0.2740 \\ 
			\hline
			30               &0.0077  &0.0480  &0.0245 &0.0353 \\
			{\tiny{ S.E.}}   &1.0135  &0.9789  &0.8957 &0.8624 \\ 
			$\hat\sigma$     &0.2486  &0.3081  &0.3052 &0.3051 \\ 
			\hline
			50               &-0.0010 &0.0296  &0.0129 &0.0187\\
			{\tiny{ S.E.}}   & 1.0153 &1.0603  &1.0324 &1.0098 \\ 
			$\hat\sigma$     & 0.2490 &0.3052  &0.3046 &0.3045 \\ 
			\hline
			80               &0.0198  &0.0244  &0.0232 &0.0244 \\
			{\tiny{ S.E.}}   &1.0207  &0.9400  &0.9194 &0.9074 \\ 
			$\hat\sigma$     &0.2492  &0.2548  &0.2547 &0.2546 \\ 
			\hline
			150              &0.0073  &-0.0044 &0.0137 &0.0032 \\
			{\tiny{ S.E.}}   &1.0044  &1.0022  &0.9988 &0.9704 \\ 
			$\hat\sigma$     &0.2496  &0.2507  &0.2506 &0.2506 \\ 
			\hline\hline			
		\end{tabular}
		}
\end{center}		
{\it\caption{Simulated Recycled distributions of $\hat{\Cal{R}}^*_{n,c}$ for Model I using various random weights showing the means and Standard Errors}}
\end{table}

\begin{table}[!h]	
\begin{center}
		{\absmall
		\vskip 2pt
		\begin{tabular}{c c c c c  }	
			\hline\hline
			Sample   & {Simulated} & {Multinomial} & {Dirichlet} & {Exponential }\\
			Size  & {Dist} & {Weights} & {Weights} & {Weights }\\
			\hline
			10               &0.0281  &0.2744  &0.1707 &0.1482 \\
			{\tiny{ S.E.}}   &1.1787  &1.0445  &0.7991 &0.7225\\ 
			$\hat\sigma$     &0.2425  &0.2907  &0.2731 &0.2732 \\ 
			\hline
			30               &0.0150  &0.0867  &0.0837 &0.0872 \\
			{\tiny{ S.E.}}   &1.0321  &1.1519  &1.0668 &1.0128 \\ 
			$\hat\sigma$     &0.2485  &0.3114  &0.3095 &0.3093 \\ 
			\hline
			50               &0.0195  &-0.0112 &0.0070 &0.0036\\
			{\tiny{ S.E.}}   &1.0221  &0.9080  &0.8712 &0.8559 \\ 
			$\hat\sigma$     &0.2489  &0.3012  &0.3007 &0.3007 \\ 
			\hline
			80               &0.0109  &0.0112  &-0.0120&0.0191 \\
			{\tiny{ S.E.}}   &1.0091  &1.0614  &1.0323 &1.0349 \\ 
			$\hat\sigma$     &0.2493  &0.2536  &0.2534 &0.2535 \\ 
			\hline
			150              &0.0175  &0.0226  &0.0120 &-0.0020 \\
			{\tiny{ S.E.}}   &1.0224  &0.8787  &0.8797 &0.8756 \\ 
			$\hat\sigma$     &0.2496  &0.2520  &0.2519 &0.2520 \\ 
			\hline\hline			
			
		\end{tabular}
		}
\end{center}
{\it\caption{Simulated Recycled distributions of $\hat {\Cal{R}}^*_{n,c}$ for Model II using various random weights showing the means and Standard Errors}}
\end{table}
To further study the {\it recycling} estimation procedure, we took $\c =(1, 0)^\t$ for obtaining sampling distribution for $\theta_1$ and $\c =(0, 1)^\t$ for that of $\theta_2$ and a $95\%$ confidence interval was constructed in each case of three distributions for random weights: {\it Dirichlet}, {\it Multinomial} and {\it Exponential} distribution. In each case, $B = 10,000$ replications of such {\it recycled} simulations were executed to determine the percentage of times the true value of the parameter was contained in the interval estimate and the average length of the confidence interval  was calculated. The simulation results for Model I, corresponding to $\theta_1$ and $\theta_2$ are provided in Table 3 and Table 4, respectively. The first column in these tables indicates the sample size $n$. Column 2 to Column 4 provide the Coverage Percentages and the average length of the confidence interval (in parentheses).

\begin{table}[h!]
		{\absmall
	\begin{center}
        \vskip 2pt
		\begin{tabular}{c c c c c }
			\hline\hline
            n & Multinomial & Dirichlet &Exponential \\
            \hline
			10     &0.921  (0.301) &0.861  (0.231) &0.862  (0.231) \\
			      
			30     &0.921  (0.150) &0.902  (0.142) &0.903  (0.142) \\
			       			
			50     &0.936  (0.116) &0.929  (0.112) &0.926  (0.112) \\
			       			
			80     &0.942  (0.092) &0.938  (0.090) &0.936  (0.090) \\
			
			150    &0.941  (0.067) &0.938  (0.067) &0.936  (0.067) \\
			      \hline\hline	
		\end{tabular}
		\end{center}
	{\it\caption{Coverage percentage (average confidence interval length) for $\theta_1$ in Model I}}
\label{4.1}	
}
\end{table}

\begin{figure}[!h]
	\centering 
	\includegraphics[width=0.6\textwidth]{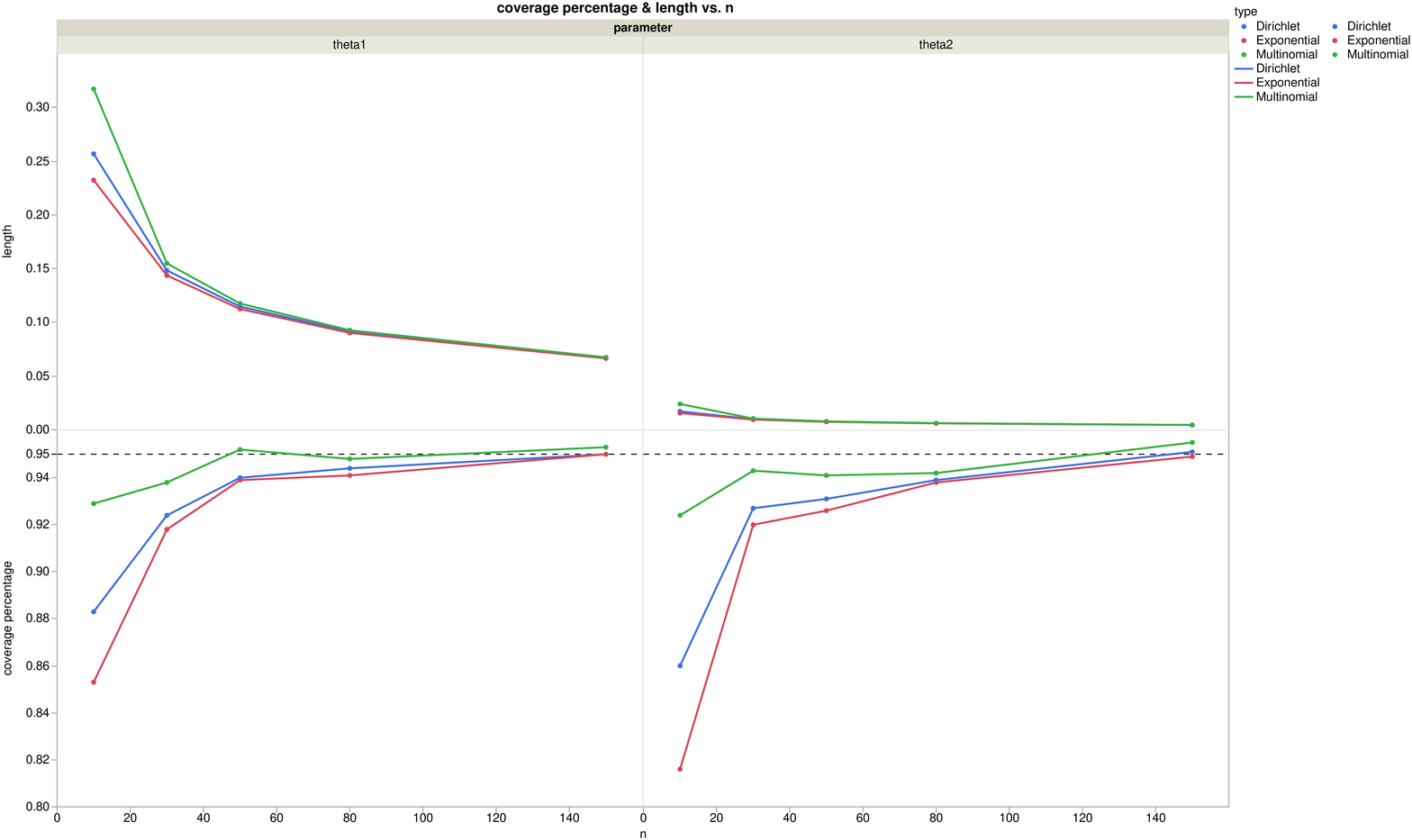}
	\caption{Coverage Percentage (average confidence interval length) for $\theta_1$, $\theta_2$ in Model I. }
	\label{6.7}
\end{figure}

\begin{figure}[!h]
	\centering 
	\includegraphics[width=0.6\textwidth]{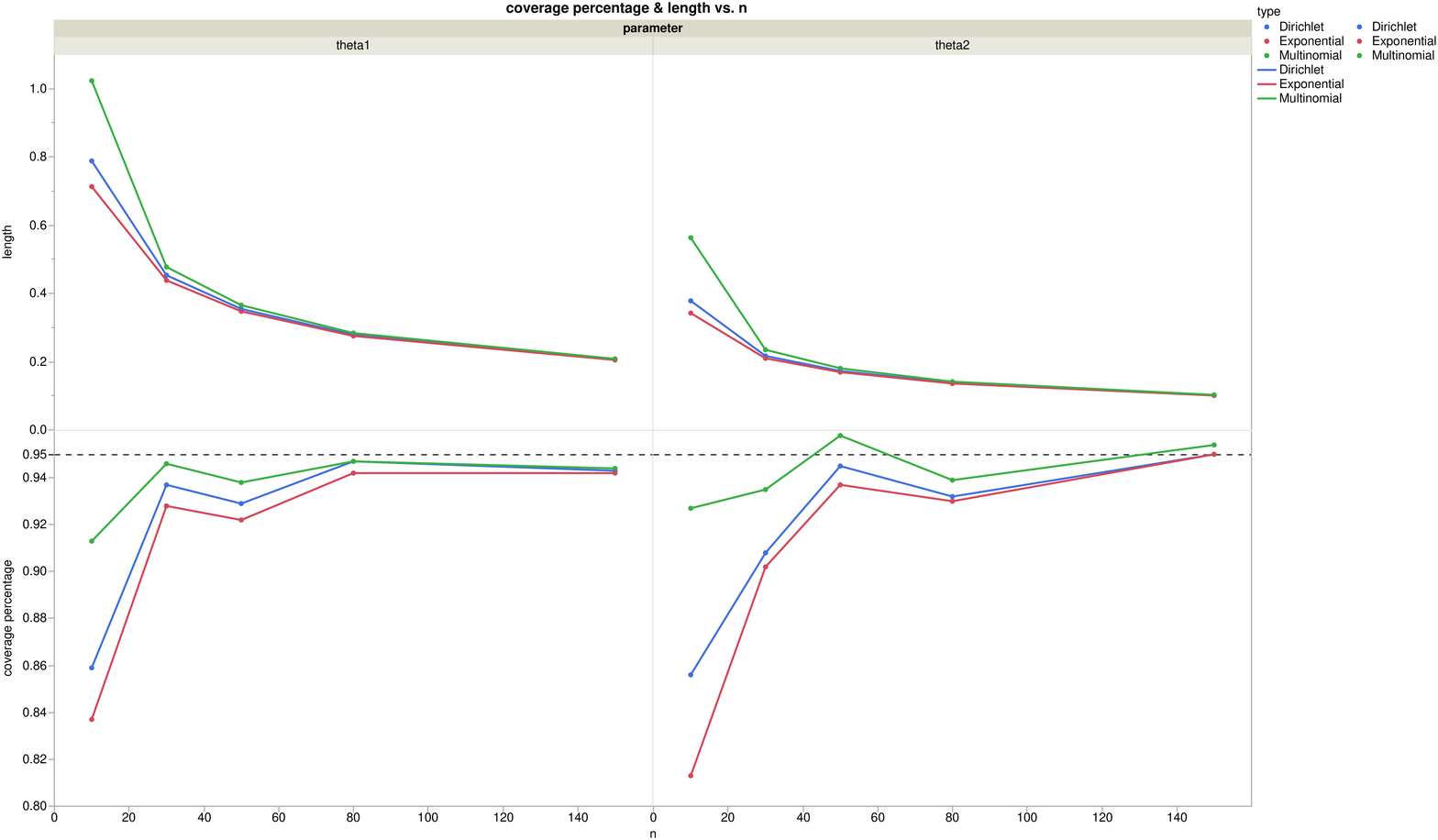}
	\caption{Coverage Percentage (average confidence interval length) for $\theta_1$, $\theta_2$ in Model II. }
	\label{6.8}
\end{figure}

\vfill\eject
\begin{table}[!h]
\label{t2}
	\begin{center}
		{\absmall
			\vskip 2pt
	\begin{tabular}{c c c c c }
		\hline\hline
		n & Multinomial & Dirichlet & Exponential \\
		\hline
		10     &0.928  (0.023) &0.841  (0.016) &0.839  (0.016)  \\
		
		30     &0.934  (0.010) &0.909  (0.009) &0.907  (0.009)  \\
			
		50     &0.932  (0.008) &0.915  (0.007) &0.915  (0.007)  \\ 
		
		80     &0.936  (0.006) &0.932  (0.006) &0.929  (0.006)  \\
		
		150    &0.951  (0.004) &0.945  (0.004) &0.943  (0.004)  \\
		\hline\hline	
	\end{tabular}\\
}
	\end{center}
	{\it\caption{ Coverage percentage (average confidence interval length) for $\theta_2$ in Model I}}
\end{table}
As can be seen from these Tables, using the {\it Multinomial} random weights for the {\it recycled} estimation performs slightly better than the other two, for both, Model I and Model II. However, the performance, in terms of the average coverage percentage,  using the {\it Dirichlet} and the {\it Exponential} weights are similar. Even when sample size is small, the confidence interval constructed by {\it recycled} estimate covers the true parameter well, especially in the case of the {\it Multinomial} random weights. We note that when the sample size increases, the Coverage Percentage for all three cases of the random weights are close to 0.95 (the nominal value), with a narrower average length  of the confidence interval. Similarly, the results of the simulation studies conducted in the case of Model II for $\theta_1$ and $\theta_2$ are illustrated in Tables 5 and 6, respectively. The coverage percentages and average confidence interval lengths for Model I and II are also illustrated in Figure \ref{6.7} and Figure \ref{6.8}.
\begin{table}[!h]
\label{4.3}	
\begin{center}
	{\absmall
	\vskip 2pt
		\begin{tabular}{c c c c c }
			\hline\hline
			n & Multinomial& Dirichlet & Exponential \\
			\hline
			10     &0.924  (0.977) &0.875  (0.716) &0.876  (0.716) \\
			
			30     &0.930  (0.468) &0.910  (0.437) &0.910  (0.437) \\
			
			50     &0.945  (0.359) &0.934  (0.345) &0.933  (0.345) \\
			
			80     &0.942  (0.285) &0.935  (0.278) &0.934  (0.278) \\
			
			150    &0.949  (0.207) &0.947  (0.205) &0.946  (0.205) \\
			\hline\hline	
		\end{tabular}
		}
    \end{center}
    {\it\caption{Coverage Percentage (average confidence interval length) for $\theta_1$ in Model II}}
\end{table}
\vfill\eject
\begin{table}[!h]
\label{4.4}
\begin{center}  
	{\absmall
	\vskip 2pt
		\begin{tabular}{c c c c c }
			\hline\hline
			n & Multinomial & Dirichlet &Exponential \\
			\hline
			10     &0.893  (0.544) &0.806  (0.355) &0.807  (0.354) \\
			
			30     &0.897  (0.517) &0.805  (0.347) &0.808  (0.347) \\
			
			50     &0.933  (0.237) &0.906  (0.215) &0.907  (0.215) \\ 
			
			80     &0.935  (0.178) &0.914  (0.168) &0.916  (0.168) \\
			
			150    &0.942  (0.139) &0.929  (0.135) &0.932  (0.135) \\
			\hline\hline	
		\end{tabular}
		}
\end{center}
{\it\caption{Coverage Percentage (average confidence interval length) for $\theta_2$ in Model II}}
\end{table}

\subsection{Illustrative Example}
 
As an illustrative example we consider  the {\tt Chwirut1} data file from NIST-nonlinear least square datasets. These data are the result of a NIST study involving ultrasonic calibration. The response variable is ultrasonic response, and the predictor variable is metal distance (Chwirut, 1979). A model was proposed as:
\[
f(x; \, \tht)= \frac{e^{-\theta_1 x}}{\theta_2+\theta_3x}
\]
where $f$ is predicted response and $x$ is metal distance. The given sample data were resampled with various  random weights ({\it Multinomial, Dirichlet} and {\it Exponential})  for a total of  $M=10,000$ times so as to produce the {\it recycled} estimate, $\ths$  of $\tht=(\theta_1, \theta_2, \theta_3)^\prime$ (and of $\sigma$ too). The results are provided in Table 7 below  which also provide the corresponding least squares estimates,  $\thn$, and $\hat \sigma$ for a comparison. 

\begin{table}[!h]
\begin{center}
{\absmall
		\vskip 2pt
		\begin{tabular}{c c c c c }
			\hline\hline
			& $\theta_1$ &$\theta_2$ &$\theta_3$ &$\sigma$ \\		
			\hline
			$\hat\theta_n$ (LSE)             & 0.1903 & 0.0061 & 0.0105 & 3.3617  \\
			\hline
			$\hat\theta^*$ (Multinomial wt.)  & 0.1907 & 0.0061 & 0.0105 & 3.4002\\ 
			S.E.          & 0.0223 & 0.0005 & 0.0009 & 0.0528  \\ 
			\hline
			$\hat\theta^*$ (Dirichlet  wt.)  & 0.1906 & 0.0061 & 0.0105 & 3.3975 \\
			S.E.          & 0.0216 & 0.0005 & 0.0009 & 0.0510 \\ 
			\hline
			$\hat\theta^*$ (Exponential wt.)  & 0.1903 & 0.0061 & 0.0106 & 3.3974\\
			S.E.          & 0.0214 & 0.0005 & 0.0009 & 0.0517 \\ 
			\hline\hline			
		\end{tabular}
		}
\end{center}
\caption{The {\it recycled}  Least Squares Estimates in the {\tt Chwirut1} example; displaying the mean and the SE of the sampling distribution based on $M=10,000$ runs, along with the estimated standard  deviation, $\sigma$.}
\end{table}

\section{Technical Details and Proofs}

This section provides the technical details and proofs of the main results of the paper stated in Theorems 1 and 2. To establish these results, we repeatedly use the results stated in Lemma \ref{l1}  below, which summaries similar technical steps from Wu (1981) (see also Corollary A there).
\begin{lemma}\label{l1}
    Let $\Theta$ be a compact subset of ${\Bbb{R}}^p$ and $g_1, g_2, \dots, g_n$ functions over $\Theta$ satisfying the following assumptions:
	\begin{itemize}
		\item[a)] $\underset{n\to\infty}{\limsup}\frac{1}{n}\sum_{i=1}^{n}\underset{\tht\in\Theta}{\sup}\, g_i(\tht)^2<\infty$
		\item[b)] and if  $\sum_{i=1}^{n}\underset{\tht\in \Theta}{\sup}\, g_i(\tht)^2=\infty$, there exists a constant $M$, independent of $i$ such that
		$$ 
		\underset{\boldsymbol{s}\ne\t; \boldsymbol{s},\t\in \Theta}{\sup}\,\left|g_i(\boldsymbol{s})-g_i(\t)\right|/|\boldsymbol{s}-\t|\le M \underset{\boldsymbol{s}\in \Theta}{\sup}\left|g_i(\boldsymbol{s})\right|<\infty
		$$
	\end{itemize}	
Then for the independent random variables $v_1, v_2, \dots, v_n$ with $E(v_i)=0$, $\underset{i}{\sup}E(v_i^2)<\infty$,  we have, uniformly over $\Theta$,   ${1\over n}\sum_{i=1}^{n}g_i(\tht)v_i\to 0\quad a.s. \ \ as\ \  n\to\infty.$	
\end{lemma}

The results in the next two Lemmas are concerned with the random weights. 
 
\begin{lemma}\label{2.4.3} With $W_{i}=(w_{i:n}-1)/\tau_n, \ i=1\dots, n$, as above, and $\bar W_n:=\frac{1}{n}\ssum W_i$ we have under {\it Assumption W},    
	\[
	\frac{1}{n}\ssum W_i\overset{p^*}{\to}0 , \ \ \ \text{as} \ n\to \infty,
	\]
	\[
	\frac{1}{n}\ssum W_i^2\overset{p^*}{\to}1 , \ \ \ \text{as} \ n\to \infty,
	\]
	and hence
	\[
	\frac{1}{n}\ssum (W_i-\bar{W}_n)^2\overset{p^*}{\to}1, \ \ \ \text{as} \ n\to \infty.
	\]
\end{lemma} 	

\noindent {\it \underbar{Proof}:} 

First note that 
\beqn
E(\frac{1}{n}\ssum W_i)^2&=&\frac{1}{n^2}\ssum E(W_i^2)+\frac{1}{n^2}\underset{i_1\ne i_2}{\sum}E(W_{i_1}W_{i_2})\\
&=&\frac{1}{n}+\frac{1}{n^2}n(n-1)O(\frac{1}{n})\to 0,
\eeqn
as $n\to \infty$ so that, 
$$
\frac{1}{n}\ssum W_i\overset{p^*}{\to}0. 
$$
Also, since by {\it Assumption W},
\beqn
E(\frac{1}{n}\ssum (W_i^2-1))^2&=&\frac{1}{n^2}\ssum E(W_i^2-1)^2+\frac{1}{n^2}\sum_{i\ne j}E(W_i^2-1)(W_j^2-1)\\
&=&\frac{1}{n}EW_1^4+\frac{1}{n}+\frac{1}{n^2}\sum_{i\ne j}EW_i^2W_j^2-\frac{1}{n^2}n(n-1)\to 0
\eeqn
as $n\to\infty$, to obtain that,
\[
\frac{1}{n}\ssum W_i^2\overset{p^*}{\to}1.
\]
Finally, we conclude that,
\beqn
\frac{1}{n}\ssum (W_i-\overline{W}_n)^2=\frac{1}{n}\ssum W_i^2-\bar{W}_n^2\overset{p^*}{\to}1
\eeqn
as $n\to \infty$. 
\qed
\begin{lemma}\label{2.4.4}
For the standardized random weights,  $W_{i}=(w_{i:n}-1)/\tau_n, \ i=1\dots, n$, of Section 3, {Assumption W.2} assures that   
$	\lim_{K\to\infty}\lim_{n\to\infty}sup\| (W_1-\overline{W}_n)I_{|W_1-\bar{W}_n|>K}\|_{_2}=0$. 
\end{lemma} 

\underline{\it Proof:} By {\it Assumption W.2} we have that  $\frac{1}{n}\sum_{i=1}^{n}(W_i-\overline{W}_n)^2\overset{p^*}{\to} 1>0$ and $\lim_{n\to\infty}E(W_1-\overline{W}_n)^4 \le  \lim_{n\to\infty}8[EW_1^4+E\overline{W}_n^4]<\infty$. Hence $(W_1-\overline{W}_n)^2$ is uniformly integrable and  finally by Chebyshev's inequality,
	\beqn
	\lim_{K\to\infty}P(|W_1-\overline{W}_n|\ge K)\le\lim_{K\to\infty}\frac{V(W_1-\overline{W}_n)}{K^2}=\lim_{K\to\infty}\frac{1}{K^2}E(W_1-\overline{W}_n)^2=0.
	\eeqn 

\qed

\begin{lemma}\label{l2}
    Let $W_{i}=(w_{i:n}-1)/\tau_n, \ i=1\dots, n$, be the standardized random weights as are given in {\it Assumption W} and  let 
$$
\xi_n\equiv \xi_n(\tht)=\frac{\tau_n}{n}\sum_{i=1}^{n}W_{i}(y_i-f_i(\tht))^2.
$$
Then, as $n\to \infty$, $\xi_n(\tht)\overset{p^*}{\to} 0$ uniformly in $\theta\in \Theta$.
\end{lemma} 
\underline{\it Proof:} We begin by rewriting $\xi_n$ as 
\beqn
	\xi_n &=&\frac{\tau_n}{n}\sum_{i=1}^{n}W_i(\epsilon_i+f_i(\thz)-f_i(\tht))^2\\
	&=&\frac{\tau_n}{n}\sum_{i=1}^{n}W_i\epsilon_i^2+\frac{\tau_n}{n}\sum_{i=1}^{n}W_id_i^2(\tht)-\frac{2\tau_n}{n}\sum_{i=1}^{n}W_i\epsilon_id_i(\tht))\\
        & \equiv &\xi_{1:n}+ \xi_{2:n}+\xi_{3:n}
\eeqn	
For each of these terms, we obtain using Jensen's inequality,
	\beqn
	E^*(\xi_{1:n})^2&=&\frac{\tau_n^2}{n^2}(\sum_{i=1}^{n}\epsilon_i^4+O(\frac{1}{n})\sum_{i<j}2\epsilon_i^2\epsilon_j^2)\\
	&\le&\frac{\tau_n^2}{n}(1+(n-1)O(\frac{1}{n}))\frac{\sum_{i=1}^{n}\epsilon_i^4}{n}\to 0 \ \ uniformly \ \ as \ \ n\to \infty
	\eeqn
	\beqn
	E^*(\xi_{2:n})^2&\le&\frac{\tau_n^2}{n^2}[\sum_{i=1}^{n}d_i^4(\tht)+O(\frac{1}{n})\sum_{i<j}2d_i^2(\tht)d_j^2(\tht)]\\
	&\le&\frac{\tau_n^2}{n^2}(1+(n-1)O(\frac{1}{n}))\sum_{i=1}^{n}d_i^4(\tht)\to0  \ \ uniformly \ \ as \ \ n\to \infty
	\eeqn
	\beqn
	E^*(\xi_{3:n})^2&=&\frac{4\tau_n^2}{n^2}[\sum_{i=1}^{n}\epsilon_i^2d_i^2(\tht)+O(\frac{1}{n})\sum_{i<j}2\epsilon_i\epsilon_jd_i(\tht)d_j(\tht)]\\
	&\le&\frac{4\tau_n^2}{n}(1+(n-1)O(\frac{1}{n}))[\frac{1}{n}\sum_{i=1}^{n}(\epsilon_i^2-\sigma^2)d_i^2(\tht)+\frac{\sigma^2}{n}\sum_{i=1}^{n}d_i^2(\tht)],
	\eeqn
	where $E^*(\xi_{3:n})^2$ converges to 0 uniformly by Lemma 1 and \noindent {\it Condition J}.\\
	Accordingly, we have $E^*(\xi_{1:n})^2$, $E^*(\xi_{2:n})^2$, $E^*(\xi_{3:n})^2$ all converge to 0 uniformly. Further,  by Chebyshev's inequality, for any $\epsilon>0$
	\beqn
	P^*(|\xi_{2:n}|\ge \epsilon)\le \frac{E^*(\xi_{2:n})^2}{\epsilon}\to0  \ \ uniformly \ \ as \ \ n\to \infty, 
	\eeqn
	so that have $\xi_{2:n}\overset{p^*}{\to}0$ uniformly. Similarly, we show that 
	 $\xi_{1:n}$, $\xi_{3:n}$ converge to 0 uniformly and in probability ($P^*$).\qed
	 
	\medskip

\underline{\it Proof of Theorem 1:} It is easy to verify that $Q_n^*(\tht):=\sum_{i=1}^{n} w_{i:n}(y_i-f_i(\tht))^2$ may be written as $Q_n^*(\tht)=Q_n(\tht) +n \xi_n$. Since  by  Jennrich(1969) (and under {\it Condition J}), $\frac{1}{n}Q_n(\tht)\to D(\tht,\thz)+\sigma^2$, uniformly,  it follows from Lemma \ref{l2} immediately that $\frac{1}{n}Q^*_n(\tht)\overset{p^*}{\to} D(\tht,\thz)+\sigma^2$ uniformly. Now let $ \tilde\tht $ be a limit point of $ \ths $. Then there is a subsequence $\hat\tht_{n_t}^*$ of $\ths$ such that $\hat\tht_{n_t}^*\to\tilde\tht$ as $ t\to \infty $. 
Hence, $ \frac{1}{n_t}Q_{n_t}^*(\hat\tht_{n_t}^*)\overset{p^*}{\to} D(\tilde\tht,\thz)+\sigma^2 $. Further, since by Lemma \ref{l2}  $ \frac{1}{n_t}Q_{n_t}^*(\hat\tht_{n_t}^*)\le \frac{1}{n_t}Q_{n_t}^*(\thz)=\frac{1}{n_t}\sum_{i=1}^{n_t}w_i\epsilon_i^2=\frac{\tau_{n_t}}{n_t}\sum_{i=1}^{n_t}W_i\epsilon_i^2+\frac{1}{n_t}\sum_{i=1}^{n_t}\epsilon_i^2 \overset{p^*}{\to} \sigma^2$. Finally, letting $ t\to\infty $, we have  $D(\tilde\tht,\thz)+\sigma^2\le\sigma^2$, so that  $ D(\tilde\tht,\thz)=0 $ and therefore $ \tilde\tht=\tht_0$ by {\it Condition J}.\qed 
\medskip
\begin{lemma}\label{l3}
	Under the conditions of {\it Assumption A},
	\[
	n^{-1}\sum_{i=1}^{n}\nabla\bphi_{i}(\tth)\to \BSigma_0^{-1} \ \ a.s.
	\]
	as $n\to \infty$, where $\nabla\bphi_{i}(\tht)$ as in (\ref{d}) and $\tth$ is any sequence such that $\tth\to\thz$ a.s..
\end{lemma}
\noindent {\it \underbar{Proof}:} 
 It is straightforward to see that 
\beqn
\frac{1}{n}\sum_{i=1}^{n}\nabla\bphi_{i}(\tth)&=& \frac{1}{n}\ssum \l \nabla f_i(\tth) \nabla f_i(\tth)^\t -(y_i-f_i(\tth)) \nabla^2f_i(\tth)\r\\
&=&\frac{1}{n}\sum_{i=1}^{n}\nabla f_i(\tth) \nabla f_i(\tth)^\t-\frac{1}{n}\sum_{i=1}^{n}\epsilon_i\nabla^2f_i(\tth)\\
&-&\frac{1}{n}\sum_{i=1}^{n}[f_i(\thz)-f_i(\tth)]\nabla^2f_i(\tth).
\eeqn
Under {\it Assumption A.2} , and since $\tth\to \thz$, a.s., it immediately follows that 
$$ 
n^{-1}\sum_{i=1}^{n}\nabla f_i(\tth)\nabla f_i(\tth)^\t\to \BSigma_0^{-1}.
$$ 
Further, under {\it Assumption A.4-A.5}, and Lemma \ref{l1}, it also follows that 
$$
n^{-1}\sum_{i=1}^{n}\epsilon_i\nabla^2 f_i(\tht)\to \b0, \ \ a.s. \ \ \text{as} \ n\to\infty,  
$$ 
uniformly in $\tht\in \Theta_\delta$. Finally, 
$$
\frac{1}{n}\sum_{i=1}^{n}[f_i(\thz)-f_i(\tth)]\nabla^2f_i(\tth)\to \b0, \ \ a.s. \ \ \text{as} \ n\to\infty, 
$$ 
upon using the Cauchy-Schwarz inequality in conjunction with  {\it Assumption A.3-A.4}.
\qed

\begin{lemma}\label{7.2.5}
	Under the conditions of {\it Assumptions A} and {\it B},
	\[
	n^{-1}\tau_n\sum_{i=1}^{n}W_i\nabla\phi_{i}(\tth)\to \b0, 
	\]
	in probability, where $W_i$ as in {\it Assumption W} and $\nabla \bphi_{i}(\tht)$ as in (\ref{d}) and $\tth$ is any sequence such that $\tth\to\thz$ a.s., as $n\to\infty$. 
\end{lemma}
{\it \underbar{Proof}:} \\
It can be verify that since  $(j,k)$th entry of the random matrix in $\frac{\tau_n}{n}\sum_{i=1}^{n}W_i\nabla \bphi_{i}(\tth)$ is
\beqn
n^{-1}\tau_n\sum_{i=1}^{n}W_i\nabla\phi_{ijk}(\tth):=n^{-1}\tau_n\sum_{i=1}^{n}W_i\left(h_{i:j}(\tth)h_{i:k}(\tth)-(y_i-f_i(\tth))a_{i:jk}(\tth)\right).
\eeqn
We obtain that, 
\beqn
&& E^*\left[n^{-1}\tau_n\sum_{i=1}^{n}W_i\nabla\phi_{ijk}(\tth)\right]^2\\
&=&n^{-2}\tau_n^2E^*\left[\sum_{i=1}^{n}W_i^2\nabla\phi_{ijk}(\tth)^2+\sum_{i\ne h}W_iW_h\nabla\phi_{ijk}(\tth)\nabla\phi_{hjk}(\tth)\right]\\
&\le&n^{-2}\tau_n^2\l(n-1)O(\frac{1}{n})+1\r\sum_{i=1}^{n}\nabla\phi_{ijk}(\tth)^2.
\eeqn
Since, by {\it Assumption W}, we have that $\tau_n^2=o(n)$, we only need to show 
 \[
 \underset{n\to \infty}{\lim}n^{-1}\sum_{i=1}^{n}\nabla\phi_{ijk}(\tth)^2< \infty.
 \] 
However with straightforward rearrangements, we have 
\beqn
n^{-1}\sum_{i=1}^{n}\nabla\phi_{ijk}(\tth)^2&=&\frac{1}{n}\sum_{i=1}^{n}\left(h_{i:j}(\tth)h_{i:k}(\tth)-(y_i-f_i(\tth))a_{i:jk}(\tth)\right)^2\\
&=&\frac{1}{n}\sum_{i=1}^{n}\left(h_{i:j}(\tth)\right)^2\left(h_{i:k}(\tth)\right)^2+\frac{1}{n}\sum_{i=1}^{n}(y_i-f_i(\tth))^2\left(a_{i:jk}(\tth)\right)^2\\
&-&\frac{2}{n}\sum_{i=1}^{n}h_{i:j}(\tth)h_{i:k}(\tth)(y_i-f_i(\tth))a_{i:jk}(\tth)\\
&=&\frac{1}{n}\sum_{i=1}^{n}\left(h_{i:j}(\tth)\right)^2\left(h_{i:k}(\tth)\right)^2+\frac{1}{n}\sum_{i=1}^{n}(y_i-f_i(\tht_0))^2\left(a_{i:jk}(\tth)\right)^2\\
&+&\frac{1}{n}\sum_{i=1}^{n}(f_i(\tht_0)-f_i(\tth))^2\left(a_{i:jk}(\tth)\right)^2 \\
&+&\frac{2}{n}\sum_{i=1}^{n}(y_i-f_i(\tht_0))(f_i(\tht_0)-f_i(\tth))\left(a_{i:jk}(\tth)\right)^2\\
&-&\frac{2}{n}\sum_{i=1}^{n}h_{i:j}(\tth)h_{i:k}(\tth)(y_i-f_i(\tth))a_{i:jk}(\tth).
\eeqn
Then, under {\it Assumption A}, {\it Assumption B} and application of Lemma \ref{l1},  and upon repeated applications of the Cauchy-Schwarz inequality, it follows that 
$$
\underset{n\to \infty}{\lim}n^{-1}\sum_{i=1}^{n}\nabla\phi_{ijk}(\tth)^2< \infty, 
$$
as required. 
\qed
\begin{lemma}\label{7.2.4}
	Under the conditions of {\it Assumptions A and B} ,
	\[
    \frac{1}{n}\ssum\bphi_{i}(\thn)\bphi_{i}(\thn)^t\to \sigma^2\BSigma_0^{-1}.	
	\]	
\end{lemma}	
{\it \underbar{Proof}:} 

Rewrite as $\bphi_{i}(\tht)$ as, 
\beqn
\bphi_{i}(\tht)&=&-(y_i-f_i(\tht))\nabla f_i(\tht)\\
&=&-\l\e_i+f_i(\thz)-f_i(\tht)\r\nabla f_i(\tht).
\eeqn
Then we have, 
\beqn
\frac{1}{n}\ssum\bphi_{i}(\thn)\bphi_{i}(\thn)^\t&=&\frac{1}{n}\ssum \epsilon_i^2\nabla f_i(\thn)\nabla f_i(\thn)^\t\\
&+&\frac{1}{n}\ssum \l f_i(\thz)-f_i(\thn)\r^2\nabla f_i(\thn)\nabla f_i(\thn)^\t\\
&+&\frac{2}{n}\ssum \epsilon_i\l f_i(\thz)-f_i(\thn)\r\nabla f_i(\thn)\nabla f_i(\thn)^\t\\
& \equiv &A_1 +A_2 +A_3.
\eeqn
Upon using the Cauchy-Schwarz inequality in conjunction with {\it Assumption B.1}, it follows that $A_2\to \b0\ \ a.s.$ and $A_3\to \b0\ \ a.s.$, as $n\to \infty$. Finally, by decomposing the term $A_1$ as
\beqn
A_1&=&\frac{1}{n}\ssum (\epsilon_i^2-\sigma^2)\nabla f_i(\thn)\nabla f_i(\thn)^\t+\frac{\sigma^2}{n}\ssum\nabla f_i(\thn)\nabla f_i(\thn)^\t\equiv A_{11}+A_{12},
\eeqn
it is easy to see that $A_{12}\to \sigma^2\BSigma_0^{-1}$, by {\it Assumption A.3}, and whereas the first term $A_{11}\to \b0$,  by using again the Cauchy-Schwarz inequality in conjunction with {\it Assumption B.1}. \\
\qed
\begin{lemma}\label{7.2.6}
	Let $\thn$ be the LSE for $\thz$. Under the conditions of {\it Assumptions A} and {\it Assumption B}  we have as $n\to\infty$, 
	\[
	-\frac{1}{\sqrt{n}}\sum_{i=1}^{n}\c^\t\l\frac{1}{n}\sum_{i=1}^{n}\nabla\bphi_i(\thn)\r^{-1} W_i\bphi_{i}(\thn)\Rightarrow {\cal N}_p(\boldsymbol{0},\sigma^2\c^\t\BSigma_0\c),
	\]
	as well as,
	\[
	-\frac{1}{\sqrt{n}}\sum_{i=1}^{n} \c^\t W_i\bphi_{i}(\thn)\Rightarrow {\cal N}_p(\boldsymbol{0},\c^\t\BSigma_0^{-1}\c).
	\]
\end{lemma}	
{\it \underbar{Proof}:}\\
Let  
\[
a_{ni}=\c^\t\l\frac{1}{n}\sum_{i=1}^{n}\nabla \bphi_i(\thn)\r^{-1}\bphi_{i}(\thn).
\]
Then clearly, 
\[
\bar{a}_{n}:=\frac{1}{n}\ssum a_{ni}=\c^\t\l\frac{1}{n}\sum_{i=1}^{n}\nabla \bphi_i(\thn)\r^{-1}\frac{1}{n}\ssum\bphi_{i}(\thn)=0.
\]
By Lemma \ref{7.2.4} we have, 
\[
\frac{1}{n}\ssum\bphi_{i}(\thn)\bphi_{i}(\thn)^t\to \sigma^2\BSigma_0^{-1}.	
\]	
Thus, by Lemma \ref{l3}, 
\beqn
\frac{1}{n}\ssum (a_{ni}-\bar{a}_n)^2 &= & \frac{1}{n}\ssum a_{ni}^2 \\
&=& \c^\t\l\frac{1}{n}\sum_{i=1}^{n}\nabla \bphi_i(\thn)\r^{-1}\frac{1}{n}\ssum\bphi_{i}(\thn)\bphi_{i}(\thn)^\t\l\frac{1}{n}\sum_{i=1}^{n}\nabla \bphi_i(\thn)\r^{-1}\c\\
\\
&& \to \sigma^2\c^\t\BSigma_0\c. 
\eeqn
Accordingly, by Lemma 4.6 of Praestgaard and Wellner(1993), combined with Lemma \ref{2.4.3} and Lemma \ref{2.4.4}, we obtain the first assertion,
\[
-\frac{1}{\sqrt{n}}\sum_{i=1}^{n}\c^\t \l\frac{1}{n}\sum_{i=1}^{n}\nabla \bphi_i(\thn)\r^{-1}W_i\bphi_{i}(\thn)\Rightarrow {\cal N}_p(\boldsymbol{0},\sigma^2\c^\t\BSigma_0\c).
\]
As for the second assertion, we let 
\[
b_{ni}=\c^\t\bphi_{i}(\thn).
\]
Then,
\[
\bar{b}_{n}:=\frac{1}{n}\ssum b_{ni}=\c^\t\frac{1}{n}\ssum\bphi_{i}(\thn)=0.
\]
Again by Lemma \ref{7.2.4}, 
\beqn
\frac{1}{n}\ssum (b_{ni}-\bar{b}_n)^2&=&\frac{1}{n}\ssum b^2_{ni}\\
&=&\c^\t\frac{1}{n}\ssum\bphi_{i}(\thn)\bphi_{i}(\thn)^\t\c\to \sigma^2\c^\t\BSigma_0^{-1}\c .
\eeqn
Finally, again by Lemma 4.6 of Praestgaard and Wellner(1993)), combined with Lemma \ref{2.4.3} and Lemma \ref{2.4.4} below, we obtain that,
\[
-\frac{1}{\sqrt{n}}\sum_{i=1}^{n}\c^\t W_i\bphi_{i}(\thn)\Rightarrow {\cal N}_p(\boldsymbol{0},\sigma^2\c^\t\BSigma_0^{-1}\c).
\]
\qed
\vfill\eject
We are now ready to present the proof of the main results stated in Theorem 2.
\bigskip

\underline{\it Proof of Theorem 2:} We denote by $S_n(\tht):= \nabla Q_n(\tht)$ and  $S^*_n(\tht):= \nabla Q^*_n(\tht)$ the gradients of $Q_n(\tht)$  and $Q^*_n(\tht)$ in (\ref{2}) and (\ref{10}), respectively, so that 
\[
S_n(\tht)=2 \sum_{i=1}^{n}\bphi_{i}(\tht),
\] and
\[
S^*_n(\tht)=2\sum_{i=1}^{n}w_{i:n}\bphi_{i}(\tht),
\] 
where as given in (\ref{a}), 
$$
\bphi_{i}(\tht)=-(y_i-f_i(\tht))\nabla f_i(\tht), \ \ i=1\dots, n.
$$
By the mean-value theorem, there exists a $ \lambda_n \in[0,1]$ such that
\be\label{c}
S^*_n(\thn)=S_n^{*}(\ths)+\nabla S_n^{*} (\tth)\cdot(\thn-\ths),
\ee
where $ \tth=(1-\lambda_n)\ths+\lambda_n\thn$. 
Since  $\thn$ and $\ths$ are such that  $S_n(\thn)= \boldsymbol{0}$ and $S_n^*(\ths)=\boldsymbol{0}$, the equation in (\ref{c}) may be re-written as 
\[
\sum_{i=1}^{n}w_{i:n}\bphi_{i}(\thn)=\sum_{i=1}^{n}w_{i:n}\nabla \bphi_{i}(\tth)\cdot(\thn-\ths), 
\]
where $\nabla \bphi_i(\tht)$ is as defined in (\ref{d}). Since $W_{i}\equiv (w_{i:n}-1)/\tau_n$, we trivially obtain that, 
\[
\tau_n\sum_{i=1}^{n}W_i\l\sum_{i=1}^{n}w_{i:n}\nabla \bphi_{i}(\tth)\r^{-1}\bphi_{i}(\thn)=\thn-\ths. 
\]
Thus, we have
\be\label{7.6}
-\frac{1}{\sqrt{n}}\sum_{i=1}^{n}\c^\t\l\frac{1}{n}\sum_{i=1}^{n}w_{i:n}\nabla \bphi_i(\tth)\r^{-1} W_i\bphi_{i}(\thn)=\tau^{-1}_n\sqrt{n}\c^\t(\ths-\thn), 
\ee
where by Lemma \ref{l3} and Lemma \ref{7.2.5},
\be\label{7.4}
\frac{1}{n}\sum_{i=1}^{n}w_{i:n}\nabla \bphi_i(\tth)=\frac{\tau_n}{n}\sum_{i=1}^{n}W_i\nabla \bphi_{i}(\tth)+\frac{1}{n}\sum_{i=1}^{n}\nabla \bphi_{i}(\tth)\overset{p^*}{\to}\BSigma_0^{-1}.
\ee
Hence, 
\beqn
&& \tau^{-1}_n\sqrt{n}\c^t(\ths-\thn)\\
&=&-\frac{1}{\sqrt{n}}\sum_{i=1}^{n}\c^t\ll\l\frac{1}{n}\sum_{i=1}^{n}w_{i:n}\nabla \bphi_i(\tth)\r^{-1}-\l\frac{1}{n}\sum_{i=1}^{n}\nabla \bphi_i(\thn)\r^{-1}\rr W_i\bphi_{i}(\thn)\\
&-&\frac{1}{\sqrt{n}}\sum_{i=1}^{n}\c^t\l\frac{1}{n}\sum_{i=1}^{n}\nabla \bphi_i(\thn)\r^{-1}W_i\bphi_{i}(\thn).
\eeqn
By Lemma \ref{l3}, Lemma \ref{7.2.6} and (\ref{7.4}),
\[
\tau_n^{-1}\sqrt{n}\c^\t(\ths-\thn)\Rightarrow {\cal N}_p(\boldsymbol{0}, \c^\t\BSigma_0\c).
\]
Finally, when combined with {\it Assumption A.2}, we have that 
\[
{\cal{R}}_{n,c}^*\Rightarrow {\cal N}(0,\sigma^2),
\]
as $n\to \infty$.

\qed

\end{document}